\documentclass[aps,prx,reprint,superscriptaddress]{revtex4-2}

\usepackage{amssymb}
\usepackage{empheq}
\usepackage{graphicx}
\usepackage{hyperref}
\usepackage{hyphenat}
\usepackage{xcolor}
\usepackage{siunitx}

\graphicspath{{../figures/}}
\hypersetup{allcolors=black}
\hypersetup{colorlinks=true}
\hypersetup{urlcolor=blue}

\newcommand{\dd}{\text{d}}
\newcommand{\hypto}{\hyp{}to\hyp{}}
\newcommand{\hy}{\hyp{}}
\newcommand{\dsh}{\text{--}}

\setcitestyle{numbers,square,citesep={,\kern-.24em}}

\begin{document}

\title{Mott Memristors based on Field-Induced Carrier Avalanche Multiplication}

\author{Francesco Peronaci}
\affiliation{Max Planck Institute for the Physics of Complex Systems, Dresden
01187, Germany}
\author{Sara Ameli}
\affiliation{Max Planck Institute for the Physics of Complex Systems, Dresden
01187, Germany}
\author{Shintaro Takayoshi}
\affiliation{Max Planck Institute for the Physics of Complex Systems, Dresden
01187, Germany}
\affiliation{Department of Physics, Konan University, Kobe, 658-8501, Japan}
\author{Alexandra Landsman}
\affiliation{Max Planck Institute for the Physics of Complex Systems, Dresden
01187, Germany}
\affiliation{Department of Physics, Ohio State University, 191 West Woodruff
Ave, Columbus, OH 43210}
\author{Takashi Oka}
\affiliation{Max Planck Institute for the Physics of Complex Systems, Dresden
01187, Germany}
\affiliation{Institute for Solid State Physics, University of Tokyo, Kashiwa
277-8581, Japan}

\begin{abstract}
We present a theory of Mott memristors whose working principle is the
non-linear carrier avalanche multiplication in Mott insulators subject to
strong electric fields. The internal state of the memristor, which determines
its resistance, is encoded in the density of doublon and hole excitations in
the Mott insulator. In the current-voltage characteristic, insulating and
conducting states are separated by a negative\hy differential\hy resistance
region, leading to hysteretic behavior. Under oscillating voltage, the response
of a voltage-controlled, non-polar memristive system is obtained, with retarded
current and pinched hysteresis loop. As a first step towards neuromorphic
applications, we demonstrate self-sustained spiking oscillations in a circuit
with a parallel capacitor. Being based on electronic excitations only, this
memristor is up to several orders of magnitude faster than previous proposals
relying on Joule heating or ionic drift.
\end{abstract}

\maketitle


\section*{\label{sec0}Introduction}

In strongly correlated materials, many-body electronic interactions cannot be
treated as a weak perturbation. A spectacular consequence is the breakdown of
standard band theory in Mott insulators, which display a charge gap despite
having nominally partially-filled bands. Even more interesting, from both
fundamental and applied points of view, are states of matter obtained from a
Mott insulator by applied pressure or chemical doping\,\cite{ Imada1998,
Lee2006}, photo-doping\,\cite{ Iwai2003, Perfetti2006, Okamoto2007,
Okamoto2010}, or applied electric field\,\cite{ Tokura1988, Taguchi2000}.

A Mott insulator under a sufficiently large electric field eventually displays
a metallic response, a phenomenon known as dielectric breakdown. Although the
insulator\hypto metal transition may result from Joule heating\,\cite{
Fursina2009, Zimmers2013}, there is growing experimental evidence that also
purely electronic transitions can occur\,\cite{ Cario2010, Guiot2013,
Nakamura2013, Stoliar2013, Yamakawa2017, Giorgianni2019, Kalcheim2019,
Zhang2019}; see Refs.\,\cite{ Woynarovich1982, Woynarovich1982b, Oka2003,
Oka2005, Eckstein2010, Oka2012, Werner2014, Stoliar2014, Li2015, Mazza2016,
Li2017, Han2018} for theoretical investigations. Particularly in narrow-gap
Mott insulators\,\cite{ Guiot2013, Stoliar2013} the dielectric breakdown
happens via carrier avalanche multiplication, whereby the kinetic energy of
accelerated carriers is converted into excitation energy of additional
carriers. While a similar mechanism occurs also in semiconductors\,\cite{
Hirori2011}, a distinctive feature of Mott materials is the non-linearity of
the process. Indeed, non-linear response to applied fields is a fingerprint of
strongly correlated insulators, which often display multivalued $I\dsh V$
characteristic with regions of negative differential resistance
$\mathcal{R}\equiv\dd V/\dd I$ ($V,I$: voltage and current across a
two-terminal device)\,\cite{ Tokura1988, Iwasa1989, Taguchi2000, Sawano2005,
Kishida2009, Kishida2011}.

The resistance of Mott insulators may vary over several orders of magnitude
across different branches of the $I\dsh V$ curve. Owing to this resistive
switch, Mott materials are promising candidates for replacing conventional
semiconducting transistors in the field of information processing. More
specifically, in neuromorphic applications\,\cite{ Wang2020} they are proposed
to fabricate memristors\,\cite{ Chua1971, Chua1976, Chua2011, Strukov2008},
electronic devices whose resistance depends on the history of the input signal,
which are regarded as the building blocks of bio-inspired novel computing
architectures\,\cite{ Yang2013, Prezioso2015, Ielmini2018, Kendall2020,
Zhu2020}.

From a formal point of view, a voltage-controlled memristive system is defined
by its state-dependent resistance, or memristance $M(x)$ ($x$: state variable)
and by the equation of motion $\dot x = f(x,V)$. The instantaneous resistance
depends, therefore, on the past voltage. From a more empirical perspective, the
fingerprint of a memristor is a pinched hysteresis loop in the $I\dsh V$ plane
when the device is subject to a bipolar periodic signal\,\cite{ Chua1976,
Chua2011}.

Following semiconducting thin films with intertwined electronic and ionic
motion\,\cite{ Strukov2008}, diverse other solid-state platforms are being
investigated as physical realizations of memristors; in particular Mott
materials, using Joule heating to locally trigger the insulating\hypto metal
transition\,\cite{ Pickett2012, Pickett2013, Kumar2017, Kumar2017b, Kumar2020,
DelValle2020}. The time scale of these devices is set by the physical
mechanism for the resistance switch and is of the order of milliseconds for
ionic drift\,\cite{ Strukov2008, Tang2016} and of nano- to microseconds for
Joule heating\,\cite{ Pickett2013}.

In this work we present a theory of a new type of memristor made of a
narrow-gap Mott insulator, whose state variable is the density of doublon
excitations, which are the charge carriers. In stark contrast with previous
proposals, the resistance switch in this memristor is based on a purely
electronic mechanism: the field-induced non-linear carrier avalanche
multiplication. This results, in particular, in a time scale set by the doublon
decay time which is of the order of picoseconds, namely up to several orders of
magnitude faster than in previous proposals.

In the following, we illustrate the microscopic working principle in
Sec.\,\ref{sec1}, where we present a phenomenological model for the
field-induced non-linear carrier avalanche multiplication. Building on this, in
Sec.\,\ref{sec2} we introduce our model of Mott memristor, derive the static
current\hyp voltage curve, and study the d.\,c.\ transitions between insulating
and conducting states. In Sec.\,\ref{sec3} we study the a.\,c.\ response,
obtaining the typical behavior of a voltage-controlled, non-polar memristive
system; and derive the steady-state diagram. Finally, in Sec.\,\ref{sec4}, as a
first step towards neuromorphic applications, we study a circuit with a
parallel capacitor and demonstrate self-sustained current oscillations,
reminiscent of the periodic spiking activity of biological neurons.


\section{\label{sec1}Phenomenological model of field-induced carrier avalanche
multiplication in Mott Insulators}

\begin{figure}
\includegraphics[]{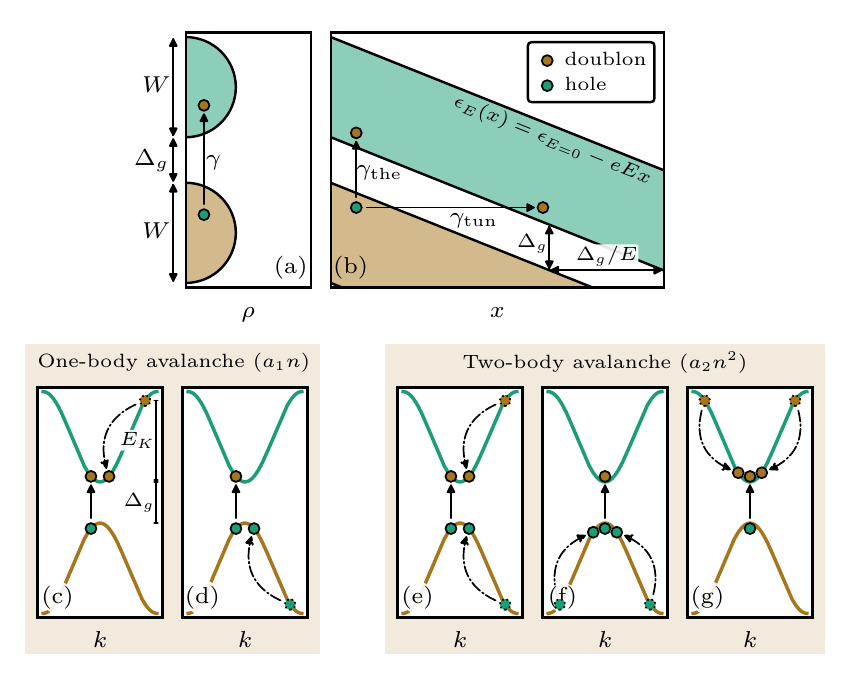}
\caption{\label{fig_0}(a) Schematic of lower and upper Hubbard bands, i.\,e.\
one-particle-excitation density of states $\rho$, in a Mott insulator with gap
$\Delta_g$ smaller than bandwidth $W$; and of doublon-hole pair excitation
($\gamma$). (b)~Band bending in real space $x$ under electric field $E$ and
doublon-hole pair creation by thermal activation ($\gamma_\text{the}$) and
quantum tunneling ($\gamma_\text{tun}$). (c)-(g)~Sketch of doublon and hole
dispersions in momentum space $k$, with one-body [(c),(d)] and two-body
[(e)-(g)] avalanche processes.}
\end{figure}

We start by presenting a phenomenological model of Mott insulator as a material
with variable concentration of charge carriers. In this model, similarly to
electrons and holes in semiconductors, the carriers are doublons and holes,
which are one-particle excitations in upper and lower Hubbard bands,
respectively, see Fig.\,\ref{fig_0}(a). Note that here we adopt a simplified
description and do not consider the dynamical nature of the Mott gap, which is
held fixed. Furthermore, we impose the doublon-hole symmetry, such that these
excitations differ only for their charge ($\pm e$) and have the same
concentration $n$, which hereafter is simply referred to as doublon density.

The density of doublon excitations~$n$ can be considered as a state variable
which determines the conductivity of the material. In this phenomenological
model, doublons have charge~$e$, effective mass~$m^*$ and they accelerate in an
electric field, before scattering after a typical time~$\tau$. This is
formalized in the Drude formula for the conductivity,
\begin{equation}
\label{eq_sigma}
\sigma(n) = e^2 (m^*)^{-1} \tau n,
\end{equation}
which relates the current density~$j$ to the electric field~$E$,
\begin{equation}
\label{eq_j}
j = \sigma(n) E.
\end{equation}
Similar forms to Eq.\,\eqref{eq_sigma} also apply to weakly correlated
materials, with for example $n$ representing the density of conduction-band
electrons. The key difference with the model at hand is in the rate equation
for~$n$, in which the strong correlations typical of Mott materials appear as a
non-linear term in the doublon density,
\begin{equation}
\label{eq_dot_n}
\dot n = \gamma - n\tau_d^{-1} + (a_1n + a_2n^2)E^2 + D\nabla^2 n.
\end{equation}
Here the source term $\gamma$ describes excitations of doublon-hole pairs
induced by thermal fluctuations or quantum tunneling across the gap, see
Fig.\,\ref{fig_0}(b). In principle, these depend on temperature and electric
field; here we hold $\gamma$ fixed and concentrate on the field dependence of
the other terms. The second term in Eq.\,\eqref{eq_dot_n} describes the decay
of doublon excitations with a typical time $\tau_d$\,\cite{ Strohmaier2010}.
The equilibrium density, namely the zero-field stationary solution, is $n =
\gamma\tau_d \equiv n_0$. The one-body avalanche term ($a_1nE^2$), also known
as impact ionization, is present in both strongly\,\cite{ Werner2014} and
weakly correlated materials\,\cite{ Hirori2011}. It describes a process in
which the kinetic energy of a carrier is converted into excitation energy of
new carriers via scattering with impurities or phonons
[Fig.\,\ref{fig_0}(c),(d)]. The two-body avalanche term ($a_2n^2E^2$), on the
other hand, describes many-body scatterings of two excitations kicking out new
carriers [Fig.\,\ref{fig_0}(e)-(g)] and is therefore proportional to the
squared carrier density. The last term describes carrier diffusion due to
density gradients; hereafter we consider the homogeneous case $\nabla^2 n = 0$.

In nonzero electric field, Eq.\,\eqref{eq_dot_n} yields two stationary doublon
densities, that is the solutions of $\dot n=0$:
\begin{equation}
\label{eq_stat_n}
\bar n(E) = \frac{n_0 \bigl[E_0^2-AE^2 \pm \sqrt{(E_0^2-AE^2)^2 -4E^2 E_0^2}
\bigr]} {2E^2}.
\end{equation}
Here $E_0\equiv(\tau_d\sqrt{a_2\gamma})^{-1}$ and $A\equiv a_1 (a_2 \gamma
\tau_d)^{-1}$ is the ratio of the one- to the two-body avalanche term for
$n=n_0$. Imposing the solutions~\eqref{eq_stat_n} to be real and positive
yields the condition $E<E_\text{th}$, with the threshold electric field
\begin{equation}
\label{eq_field_th}
E_\text{th} = E_0 \frac{\sqrt{1+A}-1}{A} \approx \frac{E_0}{2} (1-0.25A),
\end{equation}
where the approximation is valid for small $A$, namely for predominant two-body
avalanche. At this threshold, the two branches of Eq.\,\eqref{eq_stat_n} merge,
the doublon density is
\begin{equation}
\bar n(E_\text{th}) = n_0 \frac{A}{\sqrt{1+A}-1} \approx 2n_0(1+0.25A),
\end{equation}
and the current density reads
\begin{equation}
j(E_\text{th}) = \sigma( \bar n(E_\text{th}) ) E_\text{th} =
\sigma_0E_0 \equiv j_0, \quad \sigma_0 \equiv \sigma(n_0).
\end{equation}
In contrast with the threshold electric field and doublon density, the
threshold current density does not depend on the one-body constant $a_1$, but
only on the two-body constant $a_2$ (through $E_0$) and it diverges for
$a_2\rightarrow0$.

\begin{figure}
\includegraphics[width=0.49\columnwidth]{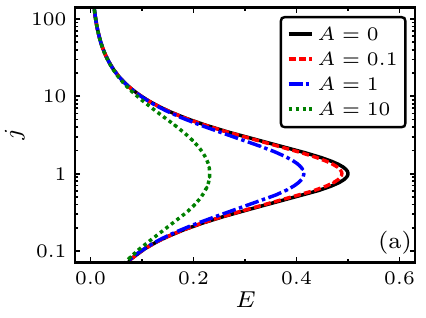}
\includegraphics[width=0.49\columnwidth]{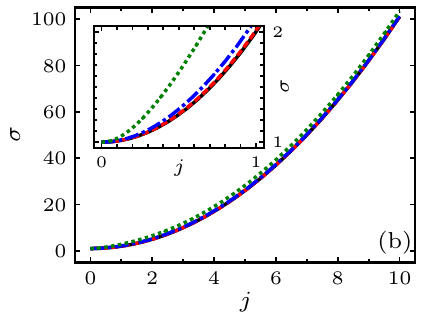}
\caption{\label{fig_1}(a)~Stationary current density versus electric field for
varying ratio of one- to two-body avalanche. (b)~Stationary conductivity
versus current density. $j_0=E_0=\sigma_0=1$.}
\end{figure}

Since the conductivity increases with doublon density, we can interpret the
lower branch of Eq.\,\eqref{eq_stat_n} as the slightly perturbed equilibrium
insulating state, and the upper branch as a conducting state. The corresponding
current density $j=\sigma(\bar n(E))E$ is plotted in Fig.\,\ref{fig_1}(a). It
should be stressed that the two branches correspond to the same microscopic
state and differ only in the doublon density; in particular, this theory does
not cover the field-induced collapse of the Mott gap.
Equation~\eqref{eq_stat_n} also implies that, within this model, there are no
stationary solutions for $E>E_\text{th}$, meaning that the material cannot
sustain such electric fields. In Fig.\,\ref{fig_1}(b) we plot the conductivity
as a function of the current density,
\begin{equation}
\begin{split}
\label{eq_stat_sigma}
\sigma(j) &= j(\bar E(j))^{-1} \\ &= \frac{\sigma_0 \bigl[j^2 +j_0^2 +
\sqrt{(j^2 +j_0^2)^2 +Aj^2 j_0^2} \bigr]} {2j_0^2} \\ &\approx \sigma_0 [1
+(j/j_0)^2 +Aj^2 (j^2 +j_0^2)^{-1}],
\end{split}
\end{equation}
where $\bar E(j)$ is the inverse function of $j(E)=\sigma(\bar n(E))E$ and the
approximation is valid for small $A$. Expressions similar to
Eq.\,\eqref{eq_stat_sigma} have been suggested to explain experiments on a
class of charge-transfer insulators\,\cite{ Tokura1988, Iwasa1989}.

The results in Fig.\,\ref{fig_1} are in qualitative agreement with experiments
in which a current is passed through a Mott insulator and the electric field
(thus the conductivity) is measured, see e.\,g.\ Refs.\,\cite{ Kishida2009,
Kishida2011}. Indeed, up to this point the treatment is suitable to describe
situations in which the current, and not the electric field, is the external
parameter. To show this from a formal point of view, we linearize
Eq.\,\eqref{eq_dot_n} around the stationary solution\,\eqref{eq_stat_n} at
fixed $E$ or at fixed $j=\sigma(\bar n(E))E$. In the former case we get $\tau_d
\delta\dot n = \pm\delta n [(1 - A(E/E_0)^{2})^2 - 4(E/E_0)^2]^{1/2}$ which
shows that only the lower branch is stable. If we instead fix~$j$, we get
$\tau_d\delta\dot n = -\delta n [1 + A(j/j_0)^2 (n_0/n)^2]$ which is stable for
all current densities. Only in the latter case states with large conductivity
are stable and can therefore be observed.

Among the parameters introduced in this section, most relevant are $\tau_d$,
$E_0$, $j_0$; which set the characteristic scales of, respectively, time,
electric field, current density. The doublon decay time is typically
$\tau_d\sim1\dsh10\,\si{\pico\second}$, as measured in ultrafast pump-probe
optical spectroscopy\,\cite{ Iwai2003, Perfetti2006, Okamoto2007, Okamoto2010},
while electric fields of the order $E_0\sim1\dsh10\,\si{\kilo\volt\per\cm}$ and
current densities $j_0\sim1\dsh10\,\si{\milli\ampere\per\cm\squared}$ have been
measured in Refs.\,\cite{ Tokura1988, Kishida2009, Kishida2011}. Together with
the physical dimensions of the memristor, $E_0$ and $j_0$ also set the
characteristic scales of, respectively, voltage and current.


\section{\label{sec2}Current-voltage characteristic and insulating-conducting
transitions}

We introduce now our model of Mott memristor as a device composed of a Mott
insulator connected in series with a conventional resistor. Adopting the
description in Sec.\,\ref{sec1}, the resistance of the Mott insulator is a
function of carrier density through the conductivity~$\sigma(n)$
[Eq.\,\eqref{eq_sigma}]:
\begin{equation}
\label{eq_resist_mott}
R(n) = LS^{-1} (\sigma(n))^{-1},
\end{equation}
where $L$ and $S$ are length and section area. Instead, the conventional
resistor has a fixed resistance $R_s$. The total resistance of the memristor,
or memristance, is therefore
\begin{equation}
\label{eq_resist}
M(n) = R(n) + R_s,
\end{equation}
and the doublon density~$n$ is its state variable. Attaching a voltage
generator~$V$ to the memristor, the electric field internal to the Mott
material is
\begin{equation}
\label{eq_field}
E = \frac{VR(n)} {L[R(n)+R_s]} = \frac{Vn_0} {L(n_0+r_sn)},
\end{equation}
with $r_s = R_s/R_0$, $R_0 \equiv R(n_0)$. Thus, the electric field does not
depend solely on the applied voltage, but also on the doublon density. For
small density the resistance of the Mott material is large, $R(n) \gg R_s$, and
the field is approximately proportional to the voltage. On the other hand, for
large density the resistance of the Mott material drops, $R(n) \ll R_s$, and so
does the field. This mechanism is crucial for the stabilization of the
conducting state of the memristor, as we discuss in this section.

The state-dependent resistance [Eq.\,\eqref{eq_resist}] and the rate equation
for the state variable [Eqs.\,\eqref{eq_dot_n} and \eqref{eq_field}] define a
non-polar voltage-controlled memristive system\,\cite{ Chua1976}. In practice,
the fixed term in the resistance corresponds to either the contact resistance,
often present especially in two-probe measurements (see e.\,g.\ Ref.\,\cite{
Sawano2005}), or a resistor added to obtain a stable conducting state\,\cite{
Tokura1988, Kishida2009, Kishida2011}.

\subsection{Stationary doublon density}

\begin{figure}
\includegraphics[]{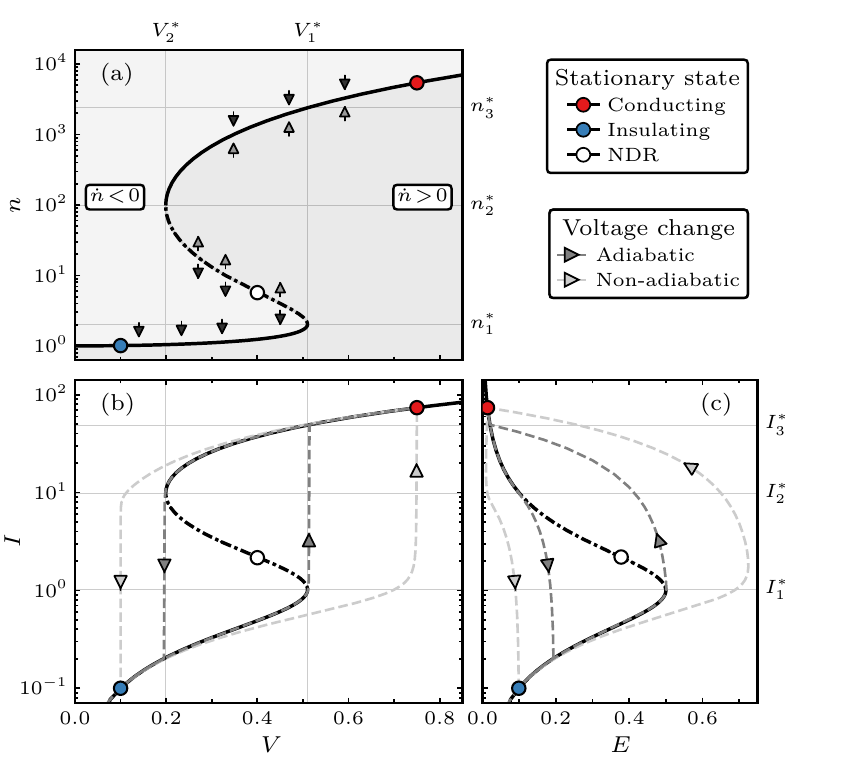}
\caption{\label{fig_2}(a) Stationary doublon density $\bar n$ versus voltage.
The arrows point up (down) where $\dot n$ is positive (negative) showing that
the solution is unstable if $\dd \bar n / \dd V <0$. (b) ``S''-shaped $I\dsh V$
curve (solid) and trajectories upon adiabatic and non-adiabatic sweep across
coexistence region $[V_2^*,V_1^*]$ (dashed). (c) Stationary current versus
internal field (solid) and same trajectories as in (b) visualized on the $I\dsh
E$ plane (dashed). $r_s=0.01$; $n_0=V_0=I_0=E_0=1$.}
\end{figure}

The stationary condition is obtained plugging Eq.\,\eqref{eq_field} into
Eq.\,\eqref{eq_dot_n} and imposing $\dot n = 0$ (we set $A=0$ hereafter). We
solve the resulting equation for $V$:
\begin{equation}
\label{eq_stat_v}
\bar V(n) = \frac{V_0(n_0+r_sn)\sqrt{n-n_0}} {n\sqrt{n_0}},
\end{equation}
where $V_0 \equiv L E_0$. This is plotted in Fig.\,\ref{fig_2}(a) as $n$ versus
$\bar V(n)$ which allows us to visualize the stationary density $\bar n$ as a
function of voltage. This solution is stable only if $\dd \bar n/ \dd V>0$,
namely for $\bar n$ outside a range $[n_1^*,n_2^*]$, where these values are
therefore obtained imposing
\begin{equation}
\label{eq_deriv}
\frac{\dd \bar V} {\dd n} = \frac{V_0(r_s n^2-n_0 n+2n_0^2)}
{2n^2\sqrt{n_0(n-n_0)}} = 0,
\end{equation}
which yields
\begin{equation}
\label{eq_nth}
n_{1,2}^* = \frac{n_0\left(1\pm\sqrt{1-8r_s}\right)}{2r_s}.
\end{equation}
For small $r_s$ we can approximate $n_1^*\approx 2n_0(1+2r_s)$ and
$n_2^*\approx n_0 r_s^{-1}$. Therefore, the two stable branches are well
separated ($n_2^*/n_1^*\approx 0.5 r_s^{-1}$) and we can interpret them as the
insulating ($n<n_1^*$) and conducting ($n>n_2^*$) states of the memristor.
Increasing $r_s$ the two branches approach each other as $n_2^*-n_1^* =
n_0\sqrt{1-8r_s}/r_s$ and eventually merge for $r_s=0.125$. Beyond this value,
we have one continuous stable state with no clear separation between insulating
and conducting states. In the opposite limit, $r_s\rightarrow0$, the stable
conducting branch vanishes ($n_2^*\rightarrow\infty$). In the remainder of this
work we set $r_s=0.01$. Between $V_2^*=\bar V(n_2^*) \approx 2V_0 \sqrt{r_s}$
and $V_1^*=\bar V(n_1^*)\approx 0.5V_0(1+2r_s)$ insulating and conducting
states coexist. In particular, to $V_1^*$ correspond the densities $n_1^*$ on
the insulating branch and $n_3^*$ on the conducting branch.

\subsection{Current-voltage characteristic}

In the stationary state with voltage $\bar V(n)$ and doublon density $n$, the
current through the memristor is
\begin{equation}
\label{eq_stat_j}
\bar I(n) = \frac{\bar V(n)} {R(n)+R_s} = \frac{I_0\sqrt{n-n_0}} {\sqrt{n_0}},
\end{equation}
where $I_0 \equiv V_0 R_0^{-1}$. Plotting Eq.\,\eqref{eq_stat_j} versus
Eq.\,\eqref{eq_stat_v} we obtain the current-voltage curve in
Fig.\,\ref{fig_2}(b). This has a distinct ``S'' shape composed of three
branches with alternating differential resistance $\mathcal{R} \equiv \dd V/\dd
I$, which is positive in the stable insulating and conducting branches; and
negative in the unstable region in between [negative-differential-resistance
region (NDR)].

A voltage sweep across the range $[V_2^*,V_1^*]$ results in a current
hysteresis, see Fig.\,\ref{fig_2}(b). If the voltage change is adiabatic,
meaning so slow that at each moment the memristor is stationary, then from the
insulating branch the current follows the $I$-$V$ curve up to $V_1^*$, where a
jump discontinuity leads from $I_1^*=\bar I(n_1^*) \approx I_0(1+2r_s)$ to the
conducting branch in $I_3^*=\bar I(n_3^*)$. Then, upon decreasing the voltage,
the current remains large down to $V_2^*$ where a second discontinuity leads
from $I_2^* = \bar I(n_2^*)\approx I_0/ \sqrt{r_s}$ back to the insulating
branch. If the voltage change is non-adiabatic, namely rapidly increasing and
decreasing, the current does not follow thoroughly the $I$-$V$ curve but
instead traces a larger hysteresis area.

In Fig.\,\ref{fig_2}(c) we plot the same quantities as in Fig.\,\ref{fig_2}(b)
versus the electric field internal to the Mott insulator. Since current and
current density are proportional, $I=jS$, the stationary curve is a rescaled
copy of Fig.\,\ref{fig_1}(a) with the crucial difference that this is now
stable also for $I>I_2^*$. The trajectories appear different in the $I\dsh E$
plane with respect to the $I\dsh V$ curves; since during the constant-voltage
insulating-conducting transitions both current and internal field vary. Also in
this case, a non-adiabatic voltage results in a wider trajectory.

\subsection{Delay time and relaxation time}

\begin{figure}
\includegraphics[width=0.49\columnwidth]{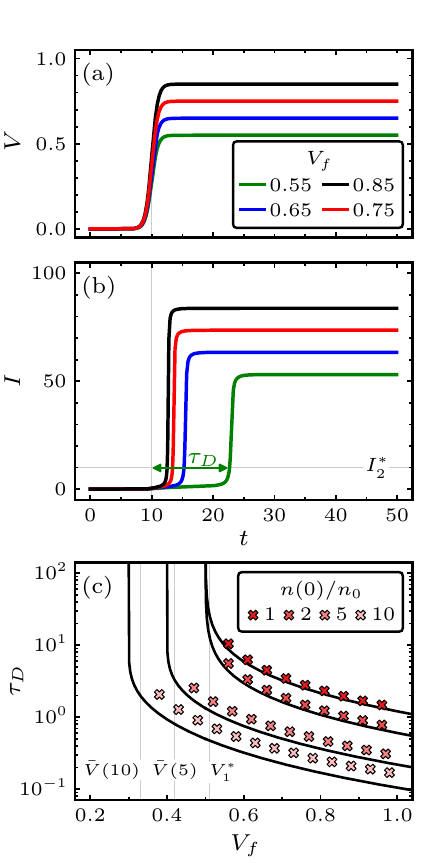}
\includegraphics[width=0.49\columnwidth]{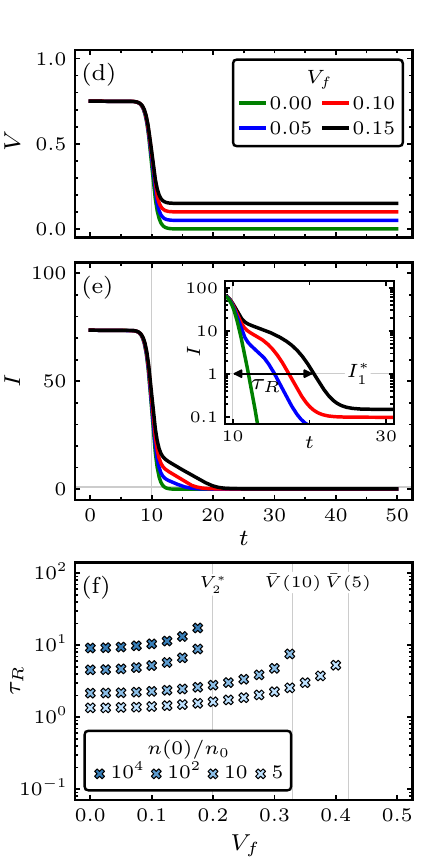}
\caption{\label{fig_3}(a) Voltage ramp to $V_f>V_1^*$ and (b) corresponding
current evolution. The delay time $\tau_D$ is the interval between the voltage
ramp and when $I=I_2^*$ (see arrow for $V_f=0.55$). (c) Delay time versus $V_f$
for various initial conditions $n(0)$ (markers) and approximation
Eq.\,\eqref{eq_delay} (solid). (d) Voltage ramp to $V_f<V_2^*$ and (e)
corresponding current evolution. The relaxation time $\tau_R$ is the interval
between the voltage ramp and when $I=I_1^*$ (see arrow for $V_f=0.15$ in the
log-scale inset). (f) Relaxation time versus $V_f$ for various initial
conditions $n(0)$. Voltage, current, time are in units of $V_0 = \SI{1}{\volt}$
(e.\,g.\ $E_0 = \SI{1}{\kilo\volt\per\cm}$, $L = \SI{10}{\micro\m}$), $I_0 =
\SI{1}{\micro\ampere}$ (e.\,g.\ $j_0 = \SI{10}{\milli\ampere\per\cm\squared}$,
$S = 100\times100\,\si{\micro\m\squared}$), $\tau_d = \SI{10}{\pico\second}$.
$r_s=0.01$.}
\end{figure}

To study the time scales associated with the transitions between insulating and
conducting states, we consider a voltage $V(t) = V_i + (V_f-V_i)f(t)$ with a
ramp function $f(t) = [1+\tanh(t-10)]/2$ and we numerically integrate
Eqs.\,\eqref{eq_dot_n},\,\eqref{eq_field}. From the insulating state, as the
voltage increases above $V_1^*$, the transition takes place in two steps
[Fig.\,\ref{fig_3}(a),(b)]: first, during a delay time $\tau_D$ the current
remains low; then, it rapidly increases above $I_2^*$, meaning that the
memristor has become conducting. Notice that after the transition
$I\propto V_f$ since in the conducting state the memristance is
approximately constant $M(n)\approx R_s$.

The delay time is plotted in Fig.\,\ref{fig_3}(c) versus the voltage and for
varying initial conditions. While the insulting\hypto conducting transition
naturally starts from the insulating branch $[n_0,n_1^*]$, here we consider
also initial conditions in the unstable region $[n_1^*,n_2^*]$ which are
relevant in the case the voltage changes while the memristor is not at
equilibrium. The delay time decreases with increasing voltage and larger
initial density. It diverges in $V_1^*$ if the initial density is below
$n_1^*\approx2.02n_0$, or in $\bar V(n(0))$ otherwise. This difference can be
explained with the aid of the stationary curve in Fig.\,\ref{fig_2}(a), which
shows that for $n>n_1^*$ the minimum voltage leading to the conducting branch
is indeed $\bar V(n)$.

To get analytical insight into the delay time and its dependence on voltage and
initial density, we solve Eq.\,\eqref{eq_dot_n} in the approximation $E\approx
VL^{-1}$ obtaining for $V>V_1^*\approx0.5V_0$ (see Appendix~\ref{appa}):
\begin{equation}
\label{eq_approx}
n(t) = \bar n_\text{av} \left[1-\Delta\cot \left[\Delta(t-\tau_D)/(2\tau_d)
\right] \right],
\end{equation}
where $\bar n_\text{av} = 2n_0 (V_1^*/V)^2$ and $\Delta = [(V/V_1^*)^2 -
1]^{1/2}$. In this approximation the transition to the conducting state
happens where Eq.\,\eqref{eq_approx} diverges, giving the delay time
\begin{equation}
\label{eq_delay}
\tau_D = (2\tau_d/\Delta)\cot^{-1} \left[ (n(0)-\bar n_\text{av})
/(\bar n_\text{av} \Delta ) \right],
\end{equation}
which we plot in Fig.\,\ref{fig_3}(c) alongside the numerical result. In the
limit $V\rightarrow (V_1^*)^+$ we have $\bar n_\text{av}\rightarrow 2n_0\approx
n_1^*$ and $\Delta\rightarrow0$. The behavior of $\tau_D$ depends on whether
$n(0)$ is smaller or larger than $n_1^*$, in the former case it diverges as
$\tau_d\approx2\tau_d\pi/\Delta$, while in the latter case it stays finite and
diverges at a lower voltage $\bar V(n(0))$.

Also the transition from the conducting state, as the voltage decreases below
$V_2^*$, takes place in various steps [Fig.\,\ref{fig_3}(d),(e)]: first, the
current rapidly decreases; then, it remains high during a relaxation time
$\tau_R$; finally, it decreases below $I_1^*$. The relaxation time is plotted
in Fig.\,\ref{fig_3}(f) versus the voltage and for varying initial conditions.
Analogously to what discussed for the delay time, we consider initial
conditions in the conducting branch $[n_1,\infty]$ as well as in the unstable
region $[n_1^*,n_2^*]$. The relaxation time increases with increasing voltage
and larger initial density; and diverges in $V_2^*$ if the initial density is
above $n_2^*\approx100\,n_0$, or in $\bar V(n(0))$ otherwise.

The results in this section, in particular the current-voltage characteristic
and the delay time, qualitatively agree with various experiments on similar
devices\,\cite{ Tokura1988, Iwasa1989, Taguchi2000, Sawano2005, Kishida2009,
Kishida2011}. Moreover, the analysis of delay and relaxation times sets the
stage for the discussion of the a.\,c.\ response, a fundamental characteristic
of a memristive system.


\begin{figure*}
\includegraphics[]{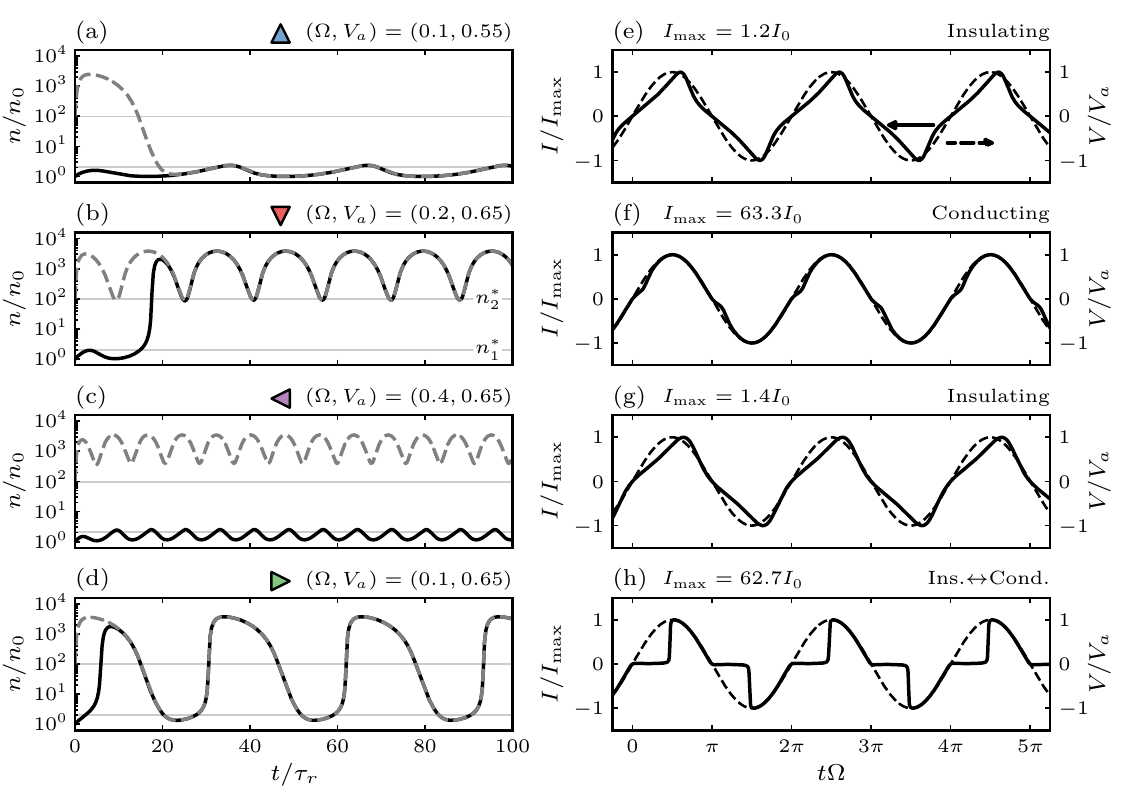}
\includegraphics[]{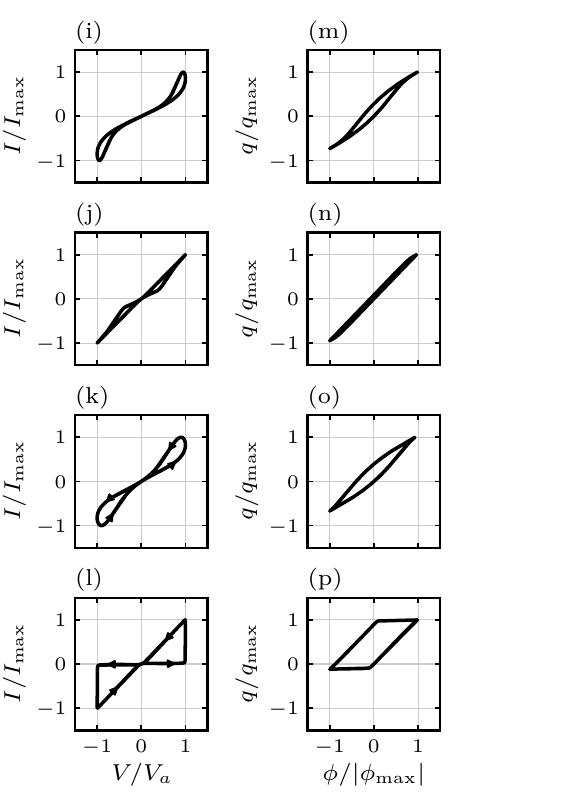}
\caption{\label{fig_4}Memristive behavior in a.\,c.\ voltage. (a)-(d) Time
evolution of doublon density with initial condition $n(0)=n_0$ (solid) or
$n(0)=100\,n_0$ (dashed) for various choices of frequency and amplitude
$(\Omega,V_a)$ (cf.\ triangle markers in Fig.\,\ref{fig_5}). The state is
insulating if $n<n_1^*$ and conducting if $n>n_2^*$. (e)-(h) Corresponding
steady-state current (solid) and applied voltage (dashed). (i)-(l) Pinched
hysteresis loop in the $I$-$V$ plane. (m)-(p) Integral of current (charge $q$)
versus integral of voltage (flux $\phi$) in the steady state. In
(g),\,(k),\,(o) only the initial condition $n(0)=n_0$ is considered.}
\end{figure*}

\section{\label{sec3}Response to alternating voltage}

We proceed now with the study of the a.\,c.\ response of the
Mott memristor introduced in Sec.\,\ref{sec2} and defined by its
state-dependent resistance and state-variable equation of motion
[Eqs.\,\eqref{eq_dot_n},\,\eqref{eq_resist_mott}-\eqref{eq_field}], including
the typical memristive features of current retardation and current-voltage
pinched hysteresis loop.

\subsection{Time evolution of doublon density and steady-state current}

In Fig.\,\ref{fig_4}(a)-(d) we plot the time evolution of doublon density
[obtained by numerical integration of Eqs.\,\eqref{eq_dot_n} and
\eqref{eq_field}] for various amplitude and frequency of the voltage $V(t) =
V_a\cos(\Omega t)$ and for two different initial conditions. We distinguish
four qualitatively different steady states. In Fig.\,\ref{fig_4}(a),(b) the
steady state is respectively insulating ($n<n_1^*$) or conducting ($n>n_2^*$)
independently of the initial condition. In contrast, in the case of
Fig.\,\ref{fig_4}(c) there are two possible steady states depending on the
initial condition. Finally, in Fig.\,\ref{fig_4}(d) the steady state goes back
and forth the insulating and conducting states.

The corresponding steady-state current is plotted in Fig.\,\ref{fig_4}(e)-(h)
for the insulating initial condition and alongside the voltage. The time axis
is rescaled with the period $T=2\pi/\Omega$ and the current and voltage axes
with their maxima, for the purpose of comparing various choices of parameters.
The insulating steady state [Fig.\,\ref{fig_4}(e),(g)] shows a clear
retardation, namely the current profile is distorted with respect to the
sinusoidal voltage. Such a retardation effect is the hallmark of memristive
systems (see e.\,g.\ Ref.\,\cite{ Strukov2008}) as it exemplifies the inertial
change of instantaneous resistance. The effect almost vanishes in the
conducting steady state [Fig.\,\ref{fig_4}(f)] because in this case the
memristance is approximately constant $M(n)\approx R_s$. Finally, in the steady
state back and forth insulating and conducting [Fig.\,\ref{fig_4}(h)] the
retardation is very pronounced; in this case the voltage effectively acts as an
adiabatic switch, as we discuss below in more detail.

\subsection{Current-voltage pinched hysteresis loop and charge-flux relation}

In Fig.\,\ref{fig_4}(i)-(l) we plot the steady-state current versus the
voltage. This curve traces a pinched hysteresis loop (so called because it
crosses the coordinate axes only in the origin) which is considered the
empirical definition of a memristive system\,\cite{Chua1976}. Also here, we
have rescaled the axes for the sake of comparing the different steady states.
The curve slope is the instantaneous inverse differential resistance
$\mathcal{R}^{-1} = \dd I/\dd V$, meaning that the greater the resistance
change, the larger the area encircled by the loop. Indeed, this is more evident
in the insulating state [Fig.\,\ref{fig_4}(i),(k)] than in the conducting state
[Fig.\,\ref{fig_4}(j)] which has almost constant resistance. In the steady
state back and forth insulating and conducting [Fig.\,\ref{fig_4}(l)] the loop
is composed of flat, vertical and steep segments. These correspond to,
respectively, insulating state, insulating\hypto conducting transition,
conducting state; while the conducting\hypto insulating transition happens near
the origin [cf.\ arrows in Fig.\,\ref{fig_4}(l)].

The direction of the loop, namely whether it is traced clockwise or anti\hyp
clockwise, is related to the polarity of the memristive system. In bipolar
memristors, e.\,g.\ based on ionic drift\,\cite{Strukov2008}, the resistance
changes depending on the sign of the input. Consequently, it is either maximum
or minimum in the origin of the $I\dsh V$ plane, and the loop is anti\hyp
clockwise for positive and clockwise for negative input. In contrast, in the
present case the memristor is non-polar, meaning the resistance change is
independent of the sign of the input, cf.\ Eq.\,\eqref{eq_dot_n}. As a result,
the loop is anti\hyp clockwise both for positive and negative inputs [see
arrows in Fig.\,\ref{fig_4}(k),(l)]. Moreover, this implies that the slope in
the origin, namely the zero-voltage inverse instantaneous resistance, is the
same for increasing or decreasing voltage.

Other characteristics of a memristive system are more conveniently discussed in
terms of the relation between charge $q$ and flux $\phi$, namely the integrals
of, respectively, current and voltage. Indeed, originally the memristance was
introduced as the quantity relating flux to charge [$\dd \phi = M(q)\,\dd q$]
similarly to how the resistance relates voltage to current [$\dd V = R(I)\,\dd
I$]\,\cite{ Chua1971}. The steady-state charge-flux relation is plotted in
Fig.\,\ref{fig_4}(m)-(p). The multivaluedness of this relation is the empirical
evidence that the memristive system belongs to the class of non-ideal
memristors\,\cite{ Chua1976}. For ideal memristors, the state-variable equation
of motion depends on the input only [$\dot x = f(V)$] giving a unique relation
between charge and flux\,\cite{ Chua1971, Strukov2008}. Instead, in the broader
class of non-ideal memristors, the equation of motion depends also on the state
variable itself [$\dot x = f(x,V)$] which yields a multivalued charge-flux
relation, as in the this case. On a practical level, an ideal memristor is
non-volatile, meaning its state does not change on zero input [$f(0)=0$], while
the state of a non-ideal memristor typically relaxes [$f(x,0)\ne0$] which makes
it a volatile memory.

\subsection{Steady-state diagram}

\begin{figure}
\includegraphics{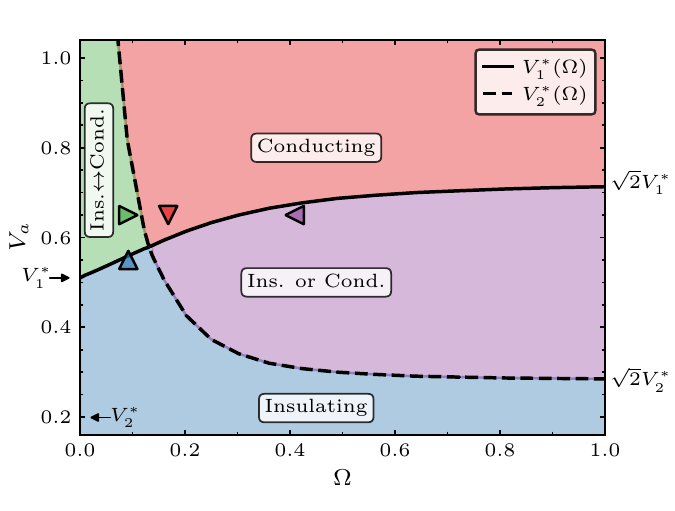}
\caption{\label{fig_5}Steady-state diagram as a function of frequency and
amplitude of a.\,c.\ voltage: insulating (blue), conducting (red), going back
and forth insulating and conducting (green) and insulating or conducting
depending on initial condition (purple). Markers indicate the four choices of
parameters in Fig.\,\ref{fig_4}. $V_{1,2}^*(\Omega)$ are the a.\,c.\ threshold
voltages for insulating-conducting transitions. Axes are in units of
$\tau_d^{-1} = 0.1\,\si{\tera\hertz}$ and $V_0 = \SI{1}{\volt}$ (e.\,g.\ $E_0 =
\SI{1}{\kilo\volt\per\cm}$, $L = \SI{10}{\micro\m}$).}
\end{figure}

In Fig.\,\ref{fig_5} we plot the steady-state diagram as a function of voltage
frequency and amplitude. This contains four regions, delimited by the
frequency-dependent amplitude thresholds $V_{1,2}^*(\Omega)$, corresponding to
each of the steady states discussed above:
\begin{enumerate}
\item For amplitude smaller than $V_{1,2}^*(\Omega)$ (blue region in
Fig.\,\ref{fig_5}) the steady state is insulating, as in Fig.\,\ref{fig_4}(a).
\item For amplitude larger than $V_{1,2}^*(\Omega)$ (red region) the steady
state is conducting, as in Fig.\,\ref{fig_4}(b).
\item For frequency not too low and amplitude within the range
$[V_2^*(\Omega),V_1^*(\Omega)]$ (purple region) the steady state is insulating
or conducting depending on the initial condition, as in Fig.\,\ref{fig_4}(c).
\item For low frequency and amplitude within the range
$[V_1^*(\Omega),V_2^*(\Omega)]$ (green region) the steady state goes back and
forth insulating and conducting, as in Fig.\,\ref{fig_4}(d).
\end{enumerate}
The a.\,c.\ thresholds $V_{1,2}^*(\Omega)$ are closely related to the d.\,c.\
thresholds $V_{1,2}^*$ of Sec.\,\ref{sec2}. If we apply a voltage with small
amplitude, such that the memristor is insulating, and then gradually increase
it, $V_1^*(\Omega)$ is the minimum value at which the memristor becomes
conducting. Notice the analogy with $V_1^*$, which is the minimum voltage to
trigger the d.\,c.\ insulating\hypto conducting transition. However, in the
a.\,c.\ case two scenarios are possible: the memristor either stays conducting
indefinitely, or it goes back to insulating at a later point of the voltage
period. The two regions above $V_1^*(\Omega)$ (respectively red and green in
Fig.\,\ref{fig_5}) correspond to these two cases. Analogously, applying a
voltage with large amplitude, such that the memristor is conducting, and then
gradually decreasing it, $V_2^*(\Omega)$ is the amplitude at which the
memristor becomes insulating.

To discuss the frequency dependence of $V_{1,2}^*(\Omega)$, it is convenient to
separately consider the regimes of low, intermediate, and high frequency.

\subsubsection{Low frequency}

At low frequency, the a.\,c.\ response to a voltage $V(t) = V_a\cos(\Omega t)$
is in a sense singular. On the one hand, at zero frequency the voltage reduces
to constant. On the other hand, at non-zero albeit low frequency, it
successively assumes all values in $[-V_a,V_a]$. In other words, in the
low-frequency limit the a.\,c.\ voltage is equivalent to an adiabatic sweep,
such as considered in Sec.\,\ref{sec2}. Thus, the steady state is insulating if
$V_a<V_1^*$ and back and forth insulating and conducting if $V_a>V_1^*$, as for
repeated sweeps, cf.\ Fig.\,\ref{fig_2}(b). Note the absence of conducting
steady states in this limit, since no matter how large the amplitude, the
memristor invariably turns insulating during the long interval in which the
voltage assumes low values. This is reflected in the divergence of
$V_2^*(\Omega)$, while $V_1^*(\Omega)$ is continuous and tends to the d.\,c.\
threshold $V_1^*$.

\subsubsection{Intermediate frequency}

The intermediate-frequency regime can be understood in terms of a competition
of time scales: the half-period $\tau_\Omega=\pi/\Omega$; and the delay
($\tau_{D}$) and relaxation ($\tau_{R}$) times, namely the time scales for,
respectively, the insulating\hypto conducting and the conducting\hypto
insulating transitions. While these were precisely defined in Sec.\,\ref{sec2}
for the d.\,c.\ transitions, here the discussion is more qualitative and
depends only on $\tau_{D,R}$ being, respectively, decreasing and increasing as
a function of voltage amplitude.

Since within the range $[V_1^*(\Omega),V_2^*(\Omega)]$ (green region in
Fig.\,\ref{fig_5}) there are one insulating\hypto conducting and one
conducting\hypto insulating transition during each half a period [cf.\
Fig.\,\ref{fig_4}(d),(h)], this region is characterized by the relation
$\tau_{D},\tau_{R}<\tau_\Omega$. Indeed, if either time scale were longer than
$\tau_\Omega$, the corresponding transition could not take place. This suggests
the interpretation of $V_{1,2}^*(\Omega)$ as the curves where, respectively,
$\tau_\Omega=\tau_D$ and $\tau_\Omega=\tau_R$. Crossing for example
$V_1^*(\Omega)$, the region with insulating steady states is characterized by
$\tau_R < \tau_\Omega < \tau_D$, that is by the inhibition of the
insulating\hypto conducting transition. Within this perspective,
$V_1^*(\Omega)$ increases with frequency because -- as $\tau_\Omega$ decreases
-- a larger voltage amplitude is needed to match the condition
$\tau_\Omega=\tau_D$.

Where $V_{1,2}^*(\Omega)$ intersect each other, the time scales are all equal:
$\tau_\Omega=\tau_D=\tau_R$. Crossing this point at constant voltage amplitude,
$\tau_\Omega$ becomes at the same time shorter than both $\tau_{D,R}$, meaning
that both the insulting\hypto conducting and the conducting\hypto insulating
transitions are inhibited, and the memristor remains in the same state as the
initial condition (purple region in Fig.\,\ref{fig_5}).

\subsubsection{High frequency}

The behavior at high frequency is better illustrated in terms of the
infinite-frequency limit, in which the voltage is equivalent to a d.\,c.\
$V_a/\sqrt{2}$. Indeed, the voltage enters the equation of motion
[Eqs.\,\eqref{eq_dot_n},\,\eqref{eq_field}] through the square $[V_a\cos(\Omega
t)]^2$ which at high frequency is equivalent to its average $V_a^2/2$. The
steady state is therefore insulating if $V_a<\sqrt{2}V_1^*$ and conducting if
$V_a>\sqrt{2}V_2^*$. Similarly to the d.\,c.\ coexistence region, in the range
$[\sqrt{2}V_2^*,\sqrt{2}V_1^*]$ the steady state is insulating or conducting
depending on the initial condition. Note that in the high-frequency limit
$V_{1,2}^*(\Omega)$ tend to constant. To reconcile this with the previous
discussion in terms of time scales, we have to consider that at high frequency
the transitions can happen across multiple voltage periods.


\section{\label{sec4}Self-sustained oscillations and spiking behavior}

We study now a first use case of the Mott memristor in electric circuits. In
the circuit in Fig.\,\ref{fig_6}(a) the memristor is connected in parallel with
a capacitor $C$ and is attached to a voltage generator $V_\ell$ through a load
resistor $R_\ell$. This setup allows us to study self-sustained current
oscillations as observed, e.\,g., in Refs.\,\cite{ Sawano2005, Kishida2009,
Kishida2011}; a phenomenon at the basis of spiking-based computational schemes.

\subsection{Nullclines and fixed point}

\begin{figure}
\includegraphics[width=0.49\columnwidth]{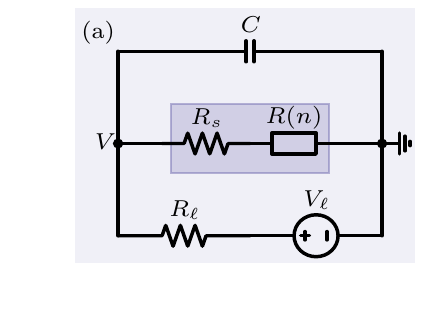}
\includegraphics[width=0.49\columnwidth]{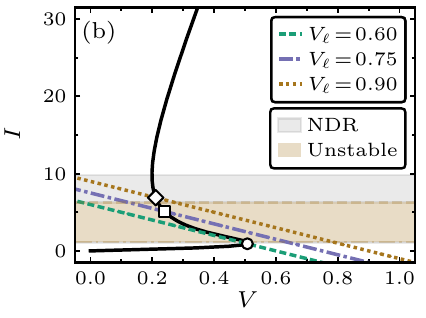}
\includegraphics[width=0.49\columnwidth]{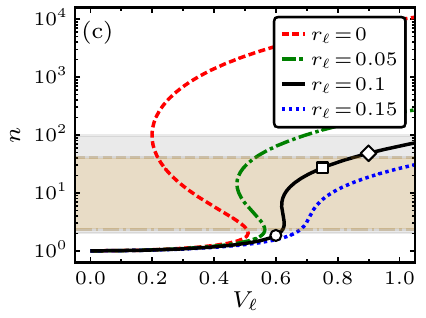}
\includegraphics[width=0.49\columnwidth]{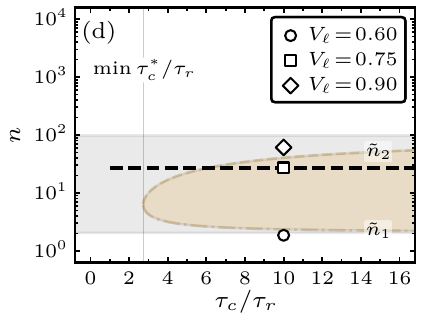}
\caption{\label{fig_6}(a)~Circuit with load voltage $V_\ell$, load resistor
$R_\ell$ and capacitor $C$ in parallel with the memristor [$R_s+R(n)$]. (b)
Fixed points at the intersection of memristor $I\dsh V$ curve (black) with load
lines $I = (V_\ell-V) R_\ell^{-1}$. (c) Fixed-point doublon density versus load
voltage. (d) Boundaries $\tilde n_{1,2}$ of the region with limit cycles
(brown) as a function of fixed-point doublon density and $\tau_c/\tau_d$; and
NDR region (gray). The same regions are highlighted in (b)-(c) for
$\tau_c/\tau_d=10$. Markers in (b)-(d) correspond to fixed points in
Fig.\,\ref{fig_7}; the dashed line in (d) to fixed points in Fig.\,\ref{fig_8}.
$n_0=V_0=I_0=1$.}
\end{figure}

The equation for the voltage $V$ across the memristor is obtained applying
Kirchhoff's law of current conservation at the nodes of the circuit in
Fig.\,\ref{fig_6}(a):
\begin{equation}
\label{eq_dot_v_2}
C\dot V + V(R_s+R(n))^{-1} + (V-V_\ell)R_\ell^{-1} = 0,
\end{equation}
which are the currents through, respectively, capacitor, memristor and voltage
generator. Equation\,\eqref{eq_dot_v_2} has to be solved together with the rate
equation for the doublon density [Eqs.\,\eqref{eq_dot_n},\,\eqref{eq_field}].
Defining $r_{\ell} = R_{\ell}/R_0$, $r_t = r_s+r_\ell$ and the time scale
$\tau_c = R_\ell C$ we rewrite these equations as a dynamical system
\begin{subequations}
\label{eq_system}
\begin{empheq}[left=\empheqlbrace]{align}
\label{eq_dot_v}
\tau_c \dot V &= V_l - V(r_tn+n_0)(r_sn+n_0)^{-1}, \\
\label{eq_dot_n_2}
\tau_d\dot n &= n_0-n + n_0n^2(V/V_0)^{2}(r_sn+n_0)^{-2}.
\end{empheq}
\end{subequations}
The fixed point of this system is at the intersection of the so-called
nullclines, namely the curves along which $\dot V=0$ and $\dot n=0$, which read
respectively
\begin{subequations}
\label{eq_nulls}
\begin{empheq}[left=\empheqlbrace]{align}
\label{eq_v_null}
V &= V_l(r_sn+n_0)(r_tn+n_0)^{-1}, \\
\label{eq_n_null}
V &= V_0(r_sn+n_0) \sqrt{n-n_0}(n \sqrt{n_0})^{-1}.
\end{empheq}
\end{subequations}
We subtract now the nullclines and, similarly to Sec.\,\ref{sec2}, we solve the
resulting equation for $V_\ell$, thereby expressing the fixed-point doublon
density as the inverse function of
\begin{equation}
\label{eq_fixed_pt}
\bar V_l(n) = \frac{V_0 (r_tn + n_0)\sqrt{n - n_0}} {n\sqrt{n_0}}.
\end{equation}
This can also be obtained imposing the intersection of the so-called load line
$I = (V_\ell-V) R_\ell^{-1}$ with the $I\dsh V$ curve of the Mott memristor
[Eqs.\,\eqref{eq_stat_v},\,\eqref{eq_stat_j}], see Fig.\,\ref{fig_6}(b), since
at the fixed point the same current flows through voltage generator and
memristor [cf.\,Eq.\eqref{eq_dot_v_2} with $C\dot V=0$].

Because the current through the capacitor is zero at the fixed point, the
resistances $R_s$ and $R_\ell$ are in series and the circuit reduces to the
situation considered in Sec.\,\ref{sec2} with the substitutions $V\rightarrow
V_\ell$ and $R_s\rightarrow R_s+R_\ell$, which indeed make
Eq.\,\eqref{eq_fixed_pt} identical to Eq.\,\eqref{eq_stat_v}. Therefore, the
same analysis applies here: if $r_t>0.125$ the solution is unique, while if
$r_t<0.125$ there is a region with three solutions, see Fig.\,\ref{fig_6}(c).
Together with $r_s=0.01$, we set hereafter $r_\ell=0.1$ which gives $r_t=0.11$.

Depending on load voltage and load resistor, the fixed point can be in the NDR
region of the Mott memristor, see Fig.\,\ref{fig_6}(b)-(d), which is necessary
for having limit-cycle self-sustained oscillations, as we discuss in the
following.

\subsection{Limit-cycle oscillations}

Self-sustained oscillations are periodic solutions of a dynamical system, such
as Eqs.\,\eqref{eq_system}, in absence of any periodic input. In the system
configuration space (here the $n\dsh V$ plane) the corresponding trajectories
are limit cycles, namely isolated closed trajectories which either attract or
repel nearby ones\,\cite{ Strogatz2018}. Simply stated, the conditions for a
limit cycle are the non-linearity of the system and the instability of its
fixed point. In this case, the former is provided by the non-linear rate
equation for the doublon density. The latter is satisfied if the fixed-point
doublon density is between the values (see Appendix\,\ref{appb})
\begin{equation}
\label{eq_region}
\tilde n_{1,2} = \frac{n_0[\tau_c - \tau_d \pm \sqrt{(\tau_c-\tau_d)^2
- 8\tau_c (r_s\tau_c+r_t\tau_d)}]}{r_s\tau_c+r_t\tau_d}.
\end{equation}
Once the load voltage -- thus the fixed-point doublon density -- is chosen,
Eq.\,\eqref{eq_region} gives an implicit expression for the critical
$\tau_c^*$, which varies with the fixed-point doublon density and whose minimum
is obtained setting to zero the argument of the square root in
Eq.\,\eqref{eq_region}:
\begin{equation}
\label{eq_min_tauc}
\min{\tau_c^*} = \frac{\tau_d[(1+4r_\ell) + \sqrt{(1+4r_\ell)^2 -
(1-8r_s)}]}{1-8r_s}.
\end{equation}
The region $[\tilde n_1,\tilde n_2]$ is included in the NDR region of the
memristor, coinciding with it in the limit of large $\tau_c$. As depicted in
Fig.\,\ref{fig_6}(d), to enter this region one can either tune the load voltage
(thus the doublon density) or the capacitor (thus the characteristic time
$\tau_c$). At this point, a supercritical Hopf bifurcation takes place\,\cite{
Strogatz2018}, namely the fixed point loses stability and a limit cycle arises.

\subsubsection{Tuning the load voltage}

\begin{figure}
\includegraphics[]{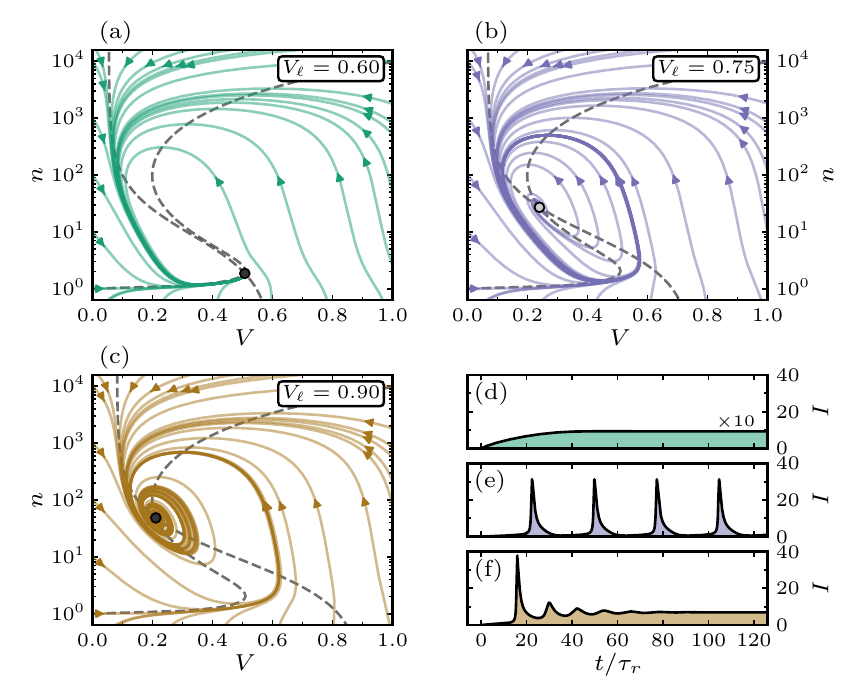}
\caption{\label{fig_7}(a)-(c) Trajectories (solid) and nullclines (dashed) in
$n\dsh V$ plane for three load voltages and $\tau_c=10$ [see markers in
Fig.\,\ref{fig_6}(b)-(d)]. The trajectories converge to the fixed point in (a)
and (c); or to a limit cycle around the fixed point in (b). (d)-(f) Current
profile for the same parameters and initial condition $(n_0,0)$ with spiking
behavior corresponding to the limit cycle (e). $r_\ell=0.1$; $n_0=V_0=I_0=1$.}
\end{figure}

At fixed $\tau_c$ we consider three load voltages such that the fixed-point
doublon density is below, inside, or above the unstable region $[\tilde
n_1,\tilde n_2]$, see Fig.\,\ref{fig_6}(b)-(d). For each load voltage we
numerically integrate Eqs.\,\eqref{eq_system} with varying initial conditions
and plot the trajectories in the $n\dsh V$ plane in Fig.\,\ref{fig_7}(a)-(c).
Outside the unstable region ($V_\ell=0.60,0.90$) all trajectories tend to the
fixed point. Notice that this implies the absence of closed trajectories. In
stark contrast, inside the unstable region ($V_\ell=0.75$) there is an isolated
closed trajectory (i.\,e.\ a limit cycle) which attracts all other
trajectories. Notice that the limit cycle is around the unstable fixed point.
In this case there is no stationary stable solution and, despite the constant
load voltage, density and voltage oscillate indefinitely. In other words, the
system undergoes limit-cycle self-sustained (or autonomous) oscillations.

The current profile is markedly different in the three cases. Let us consider
[see Fig.\,\ref{fig_7}(d)-(f)] the trajectories with initial condition
$(n_0,0)$. For $V_\ell=0.60$ the current increases monotonically to the stable
fixed point, which is on the insulating branch, see Fig.\,\ref{fig_7}(d). In
contrast, for $V_\ell=0.90$ the stable fixed point is near the conducting
branch and is reached only after a transient, which in the $n\dsh V$ plane
takes the form of a spiral around the fixed point [Fig.\,\ref{fig_7}(c)], and
the current profile has a single spike followed by damped oscillations, see
Fig.\,\ref{fig_7}(f).

Finally, corresponding to the limit cycle, for $V_\ell=0.75$ the current has
periodic spiking, see Fig.\,\ref{fig_7}(e). Each spike consists of a sudden
increase and a similarly rapid, but slower, decrease. These are due to repeated
transitions between the memristor insulating and conducting states, Note that
this is consistent with the spiking behavior of biological neurons, in which
the neural-cell membrane also transitions between insulating and conducting in
the course of an oscillation\,\cite{ Izhikevich2007}.

\subsubsection{Tuning the capacitor}

\begin{figure}
\includegraphics[]{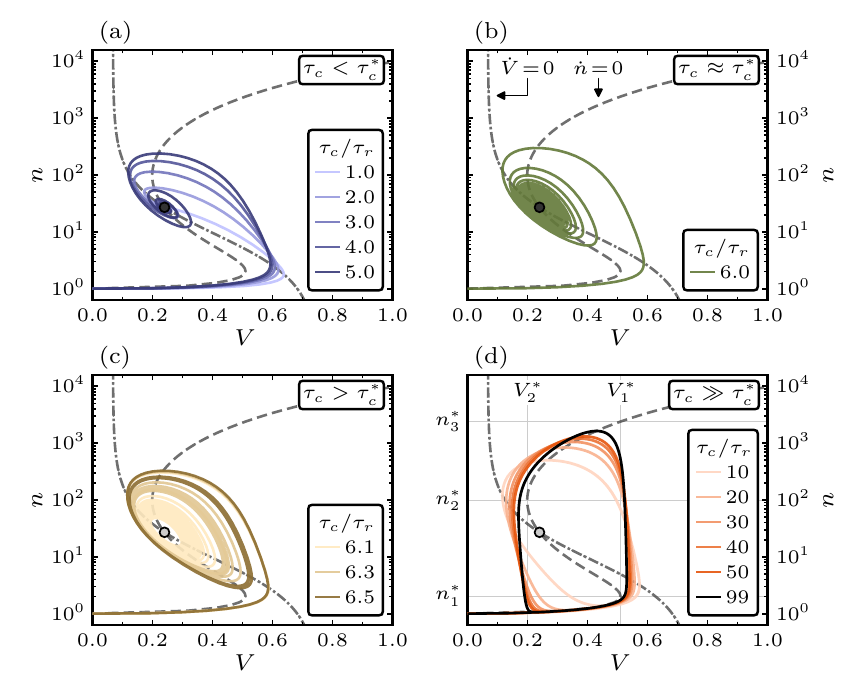}
\caption{\label{fig_8}Trajectories (solid) and nullclines (dashed) in $n\dsh V$
plane for $V_\ell=0.75$ and varying $\tau_c$ [see dashed line in
Fig.\,\ref{fig_6}(d)] with initial condition $(n_0,0)$. The fixed point is
stable in (a) and (b); and unstable in (c) and (d) where it is encircled by a
limit cycle whose area and size change with $\tau_c$.
For very large $\tau_c$ (d) the limit cycle tend to an adiabatic
hysteresis loop [cf.\ Fig.\,\ref{fig_2}(a),(b)]. $r_\ell=0.1$; $n_0=V_0=1$.}
\end{figure}

In Fig.\,\ref{fig_8} we plot the trajectories obtained by numerical solution of
Eqs.\,\eqref{eq_system} with initial condition $(n_0,0)$, fixed load voltage
and varying $\tau_c$. With its location unaltered, the fixed point loses
stability across a critical $\tau_c^*\approx6.06\,\tau_d$ (for $V_\ell=0.75$),
see Fig.\,\ref{fig_6}(d). It is stable for $\tau_c<\tau_c^*$ and reached after
a number of oscillations which become more dense as $\tau_c^*$ is approached.
As soon as $\tau_c>\tau_c^*$, the fixed point becomes unstable and a small
limit cycle appears. Increasing $\tau_c$ further, the limit cycle grows and
tends to a loop with segments at constant voltage connecting lower and upper
branches of the $\dot n=0$ nullcline, see Fig.\,\ref{fig_8}(d). Since this
nullcline is nothing but the stationary doublon density $\bar n$ versus the
voltage [cf.\ Fig.\,\ref{fig_2}(a)] this limit cycle is equivalent to the
hysteresis loop in adiabatic voltage considered in Sec.\,\ref{sec2}. In other
words, in this limit the circuit behaves like a relaxation oscillator\,\cite{
Strogatz2018}.

\begin{figure}
\includegraphics[]{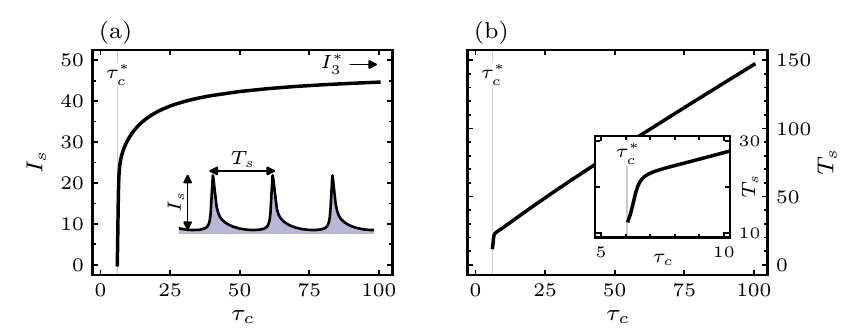}
\caption{\label{fig_9}(a) Height and (b) period of current spikes along the
limit cycle [see inset in (a)] as a function of $\tau_c$ at fixed $V_\ell=0.75$
and $r_\ell=0.1$. Current, time are in units of $I_0 =
\SI{1}{\micro\ampere}$ (e.\,g.\ $j_0 = \SI{10}{\milli\ampere\per\cm\squared}$,
$S = 100\times100\,\si{\micro\m\squared}$), $\tau_d=10\,\si{\pico\second}$.}
\end{figure}

The limit-cycle current spikes can be characterized by height (difference
between maximum and minimum) and period, see Fig.\,\ref{fig_9}. Evidently,
these quantities are only defined for $\tau_c \ge \tau_c^*$. At $\tau_c^*$ we
have the typical behavior for a supercritical Hopf bifurcation\,\cite{
Strogatz2018}: the height grows from zero (the limit cycle has vanishing
amplitude) while the period is finite and equal to $2\pi[\text{det}(J)]^{-1/2}
\approx 12\,\tau_d$, with $J$ the Jacobian of the dynamical
system\,\eqref{eq_system} at the fixed point (see Appendix\,\ref{appb}).
Increasing $\tau_c$, the height first rapidly increases, then it slowly
saturates to a value close to $I_3^*$ which is, together with $I_1^*$, the
stationary current at the threshold voltage $V_1^*$, cf.\,Fig.\,\ref{fig_2}(b).
At the same time, already at $\tau_c \approx 7\,\tau_d$, the period is linear
in $\tau_c$, showing a decoupling of time scales for doublon density and
voltage, as expected for a relaxation oscillator.


\section*{\label{sec5}Conclusions}

We have proposed the narrow-gap Mott insulator as a compact realization of a
new type of memristor based on the field-induced carrier avalanche
multiplication. Due to this purely electronic mechanism for the resistive
switch, this Mott memristor has a characteristic time scale set by the
doublon-excitation decay time $\tau_d\sim1\dsh 10\,\si{\pico\second}$, which is
up to several orders of magnitude faster than in devices based on Joule heating
or ionic drift.

As a first step we have put forward a phenomenological description of the
field-induced carrier avalanche in Mott insulators, in which the conductivity
depends on the carrier density, whose rate equation contains the non-linear
scattering terms induced by strong correlations. Building on this, we have
introduced the Mott memristor as a device made of a Mott material in series
with a conventional resistor; and we have derived its current-voltage curve, as
well as the transitions between conducting and insulating states. While the
very definition qualifies the model as a non-polar, voltage-controlled
memristive system, we have analyzed in detail its a.\,c.\ response, in
particular the pinched hysteresis loop and the steady-state diagram as a
function of amplitude and frequency. Finally, we have considered a circuit with
a capacitor in parallel with the Mott memristor, and demonstrated
self-sustained current oscillations and periodic spiking behavior, consistent
with the periodic activity of biological neurons.

While similar devices have been subject of intensive experimental study\,\cite{
Tokura1988, Iwasa1989, Taguchi2000, Sawano2005, Kishida2009, Kishida2011}, this
is the first time (to the best of our knowledge) they are proposed as
memristors. Moreover, our work provides a comprehensive theory of the key
features of those prior studies: threshold electric field, negative
differential resistance (NDR), multivalued current\hyp{}voltage characteristic,
delay time, and current oscillations. At the same time, our proposal consists
of a tractable set of equations; which stands in contrast with previous more
complicated models, see e.\,g.\ Ref.\,\cite{ Pickett2013}, and results in two
valuable features. First, it allowed us to derive analytical expressions, such
as the boundaries of the NDR region and the conditions for limit-cycle
oscillations. Second, and perhaps more importantly, it makes promising to
include the model into the description of circuits of growing complexity, in
the quest for bio-inspired novel computing architectures.


\appendix

\section{\label{appa}Derivation of Eqs.\ \eqref{eq_approx} and \eqref{eq_delay}}

In this appendix we derive Eqs.\,\eqref{eq_approx},\,\eqref{eq_delay} of
Sec.\,\ref{sec2} for doublon density and delay time of the d.\,c.\
insulating\hypto conducting transition. To simplify the exposition, we set
$\tau_d=n_0=E_0=1$, $A=0$. Then Eq.\,\eqref{eq_dot_n} is rewritten as
\begin{equation}
\label{eq_dot_n_app}
\dot n = 1 - n + n^2 E^2.
\end{equation}
The two stationary solutions are $\bar n = \bar n_\text{av}(1\pm i\Delta)$
where $\bar n_\text{av} = 1/(2E^2)$ and $\Delta = \sqrt{4E^2-1}$
[cf.\,Eq.\,\eqref{eq_stat_n}] and are real only if $E<E_\text{th}=0.5$
[cf.\,Eq.\,\eqref{eq_field_th}]. Since during the delay time the doublon
density does not change much, we approximate the field in the Mott insulator as
constant, $E\approx VL$, which is equivalent to approximating $r_s\approx 0$,
yielding $V_1^*\approx0.5V_0$ and $n_1^*\approx0.5 n_0$.
Equation\,\eqref{eq_dot_n_app} can then be solved with a variable change:
\begin{gather}
\label{eq_change}
n = -\dot x/(xE^2),\\
\label{eq_aux}
\ddot x + \dot x + E^2x = 0.
\end{gather}
The general solution of the transformed equation\,\eqref{eq_aux} is $x =
\alpha_1 e^{s_1 t} + \alpha_2 e^{s_2 t}$ where $s_{1,2} = (-1\pm i\Delta)/2$.
Substituting this back into\,\eqref{eq_change} yields the solution of
Eq.\,\eqref{eq_dot_n_app}:
\begin{equation}
\label{eq_sol_1}
n = \bar n_\text{av}\left(1-i\Delta\frac{\alpha_1 e^{i\Delta t/2}-\alpha_2
e^{-i\Delta t/2}} {\alpha_1 e^{i\Delta t/2}+\alpha_2 e^{-i\Delta t/2}}\right).
\end{equation}
Notice that the solution of \eqref{eq_aux} depends on both
$\alpha_{1,2}$ while Eq.\,\eqref{eq_sol_1} depends only on their ratio.
To proceed, we parametrize
$\alpha_{1,2}=\pm\exp(\mp i\Delta \tau_D/2)$ and obtain
\begin{gather}
\label{eq_sol_2}
n = \bar n_\text{av}\bigl[1-\Delta\cot[(\Delta/2)(t-\tau_D)]\bigr], \\
\label{eq_delay_2}
\tau_D = (2/\Delta) \cot^{-1} \bigl[ (n(0) - \bar n_\text{av})/(\Delta \bar
n_\text{av}) \bigr].
\end{gather}
which coincide with Eqs.\,\eqref{eq_approx} and \eqref{eq_delay}.

Up to now we have considered the electric field above threshold $E>E_\text{th}$,
which is equivalent to $V>V_1^*$ and makes $\Delta$ and
Eqs.\eqref{eq_sol_2},\,\eqref{eq_delay_2} real. In the limit $V\rightarrow
V_1^*$ we have $\Delta\rightarrow0$, $\bar n_\text{av}\rightarrow
2n_0\approx n_1^*$ and the behavior of the delay time Eq.\,\eqref{eq_delay_2}
depends on the initial condition:
\begin{equation}
\tau_D \approx
\begin{cases}
2\pi/\Delta, \quad & \text{if $n(0)<n_1^*$}, \\
2 n_1^*/(n(0)-n_1^*), \quad & \text{if $n(0)>n_1^*$}.
\end{cases}
\end{equation}
Indeed with a large initial density the transition happens even below
threshold. In this case we have to choose $\alpha_{1,2}$ differently or,
alternatively, we can analytically continue
Eqs.\,\eqref{eq_sol_2},\,\eqref{eq_delay_2} with $\tilde \Delta = i\Delta$
which yields
\begin{gather}
\label{eq_sol_3}
n = \bar n_\text{av} \bigl[1 - \tilde\Delta
\coth[(\tilde\Delta/2)(t-\tau_D)]\bigr], \\
\label{eq_delay_3}
\tau_D = (2/\tilde\Delta) \coth^{-1} \bigl[ (n(0) - \bar n_\text{av}) /
(\tilde\Delta \bar n_\text{av}) \bigr].
\end{gather}
In this case the delay time diverges for $n(0) = \tilde\Delta (n_\text{av}+1)$,
namely for $V=\bar V(n(0))$, as shown in Fig.\,\ref{fig_3}(c).


\section{\label{appb}Derivation of Eq.\,\eqref{eq_region}}

In this appendix we derive Eq.\,\eqref{eq_region} for the region with limit
cycle in Sec.\,\ref{sec4}. A limit cycle is guaranteed to exist by the
Poincar\'e–Bendixson theorem when the system is confined in a region with no
stable fixed point therein\,\cite{Strogatz2018}. Such a trapping region is
(with $V_\ell r_t r_s^{-1}>V_1^*$) $\{(n,V)\in[0,\bar n(V_\ell r_t r_s^{-1})]
\times [0,V_\ell r_t r_s^{-1}]\}$. The fixed point turns from stable to
unstable (Hopf bifurcation) when, with positive determinant, the trace of the
Jacobian becomes positive. For the system\,\eqref{eq_system} the Jacobian reads
\begin{equation}
\label{eq_jacobian}
J(n,V) =
\begin{pmatrix}
-\frac{r_tn+n_0}{\tau_c(r_sn+n_0)} &
-\frac{Vr_ln_0}{\tau_c(r_sn+n_0)^2} \\
\frac{2n^2n_0VV_0^{-2}}{\tau_d(r_sn+n_0)^2} &
-\frac{1}{\tau_d}+\frac{2nn_0^2(V/V_0)^{2}}{\tau_d(r_sn+n_0)^3}
\end{pmatrix}.
\end{equation}
Plugging Eq.\,\eqref{eq_n_null} for the $\dot n=0$ nullcline into
Eq.\,\eqref{eq_jacobian}, we obtain the Jacobian as a function of the
fixed-point doublon density:
\begin{equation}
\label{eq_jacobian_2}
J(n) =
\begin{pmatrix}
-\frac{r_tn+n_0}{\tau_c(r_sn+n_0)} &
-\frac{V_0r_l[n_0(n-n_0)]^{1/2}}{\tau_c n(r_sn+n_0)} \\
\frac{2n[n_0(n-n_0)]^{1/2}}{\tau_d V_0(r_sn+n_0)} &
\frac{-r_sn^2+nn_0-2n_0^2}{\tau_d n(r_sn+n_0)}
\end{pmatrix},
\end{equation}
whose determinant and trace read
\begin{gather}
\label{eq_det}
\text{det}(J) = \frac{r_tn^2-n_0n+2n_0^2} {\tau_d\tau_c(r_sn+n_0)n}, \\
\label{eq_tr}
\text{tr}(J) = \frac{\tau_c(-r_sn^2+n_0n-2n_0^2) - \tau_d(r_tn^2+n_0n)}
{\tau_d\tau_c(r_sn+n_0)n}.
\end{gather}
The sign of the determinant does not depend on $\tau_c$ and is positive for $n$
outside the range $[\hat n_1,\hat n_2]$ with $\hat n_{1,2} = n_0
(1\pm\sqrt{1-8r_t}) (2r_t)^{-1}$. The sign of the trace depends on $\tau_c$.
Notice that a necessary condition for the trace to vanish is $(-r_sn^2 +n_0n
-2n_0^2)>0$ which is the same condition for the NDR region of the memristor,
cf.\ Eq.\,\eqref{eq_deriv}, demonstrating that the region with limit-cycle
oscillations is a subset of the NDR region, as depicted in
Fig.\,\ref{fig_6}(b)-(d). Imposing the trace to be positive we get the
condition that $n$ should be outside the range $[\tilde n_1,\tilde n_2]$ with
$\tilde n_{1,2}$ given in Eq.\,\eqref{eq_region}.

\bibliography{main}

\end{document}